\documentclass[
    ,final            % use final for the camera ready runs
  ]
  {aipproc}

\layoutstyle{6x9}

\def \D0 {D\O }

\begin{document}

\title{Hard Diffractive Results and Prospects at the Tevatron\footnote{Presented at the XXXV International 
Symposium on Multiparticle Dynamics 2005, August 9-15, 2005, Kromeriz, Czech Republic}}


\author{Krisztian Peters}{
  address={Department of Physics \& Astronomy, University of
  Manchester, \\ Manchester M13 9PL, UK\\ \vspace{.3cm} {\rm on behalf
  of the \D0 ~and CDF Collaborations}} }

\begin{abstract}
We review hard diffractive results and prospects at the Tevatron with
an emphasis on factorization breaking in diffractive processes. Upper
limits on the exclusive di-jet and $\chi_c^0$ production cross
sections at CDF and the status of the \D0 ~Forward Proton Detectors are
discussed.
\end{abstract}

\maketitle

%%%%%%%%%%%%%%%%%%%%%%%%%%%%%%%%%%%%%%%%%%%%
%% MAINMATTER
%%%%%%%%%%%%%%%%%%%%%%%%%%%%%%%%%%%%%%%%%%%%
\section{Diffraction at the Tevatron}

Diffractive events are mediated by the exchange of color singlets with
vacuum quantum numbers and have clear experimental signatures. 
These are on one hand rapidity gaps: the absence of particles in some
regions of rapidity in contrast to non-diffractive events where gaps
are filled by additional soft parton interactions which yields an
exponential suppression of rapidity gaps. On the other hand tagged
protons or anti-protons: $p$ or $\bar p$ scattered at small angle and
measured in Roman Pots far away from the interaction point. Depending
on these rapidity gaps and tags, the main diffractive event topologies
at the Tevatron are: single diffraction (SD), double diffraction (DD)
or double Pomeron exchange (DPE). Single diffraction is characterized
by a leading proton or anti-proton which escapes the collision intact
and the presence of a further particle or a di-jet resulting from a
hard scattering separated by a rapidity gap. Double diffraction is
characterized by a gap in the central region and dissociated protons
and anti-protons. In DPE there is a gap on both, the proton and
anti-proton side, with a central produced di-jet or other
particles. Proton and anti-proton remains intact.

In RUN II the CDF Forward Proton Detectors (FPDs) include Roman Pot
detectors approximately 57m from the interaction point. These consist
of three stations and each station comprises one scintillation fiber
detector and one trigger counter. Beam Shower Counters, a set of
scintillation counters around the beam pipe, are used to reject
non-diffractive (ND) background at the trigger level. This makes it
possible to collect diffractive data at high luminosities. The
energy flow in the event in the very forward direction is measured by
Miniplug Calorimeters. These consist of alternating layers of lead
plates and liquid scintillator readout. It has a towerless geometry
without dead regions due to the lack of internal mechanical
boundaries. The \D0 ~FPDs are described in the following.

\section{Factorization in diffraction}

\begin{figure}[t]
(a)\includegraphics[width=.5\textwidth]{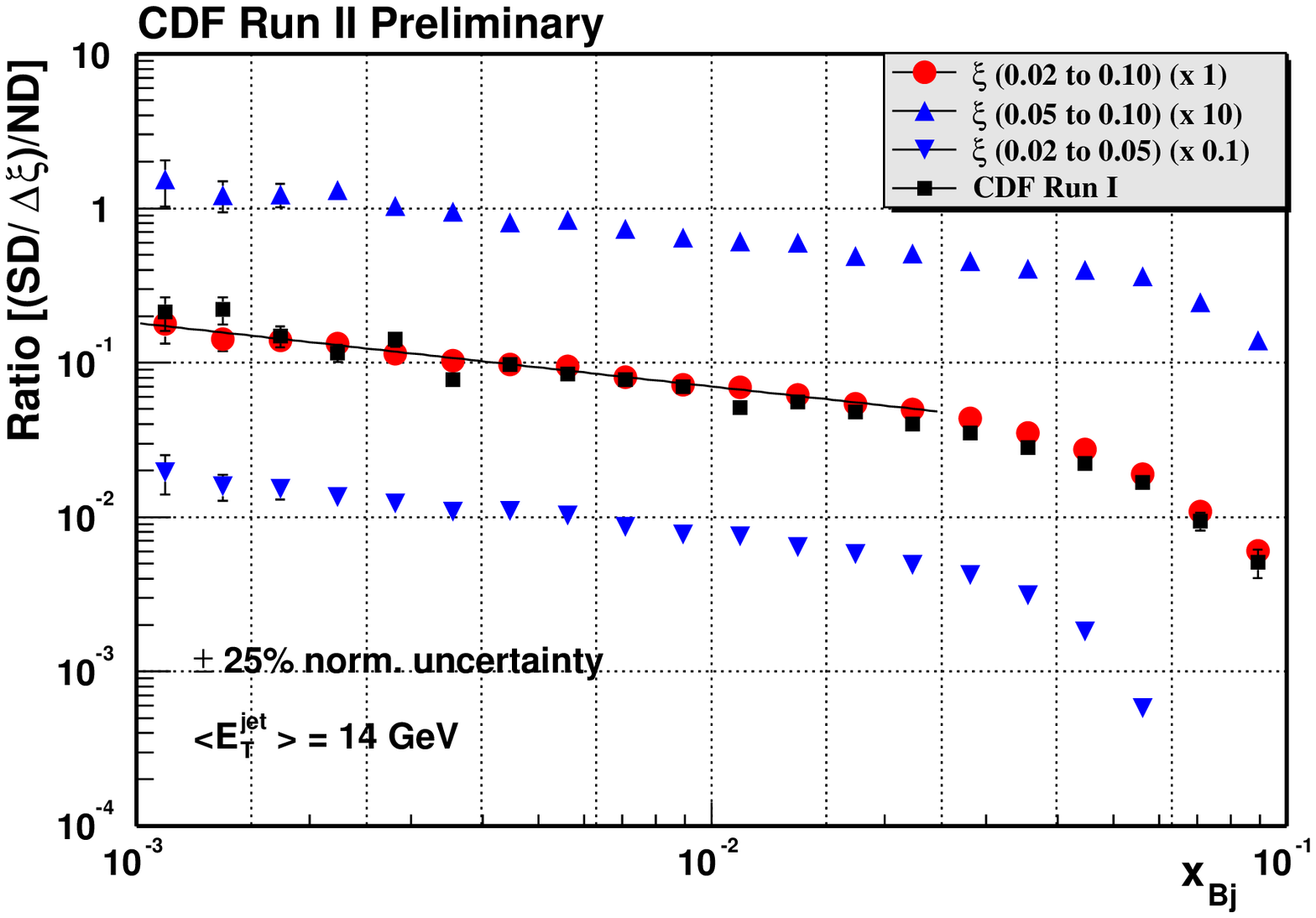}
(b)\includegraphics[width=.5\textwidth]{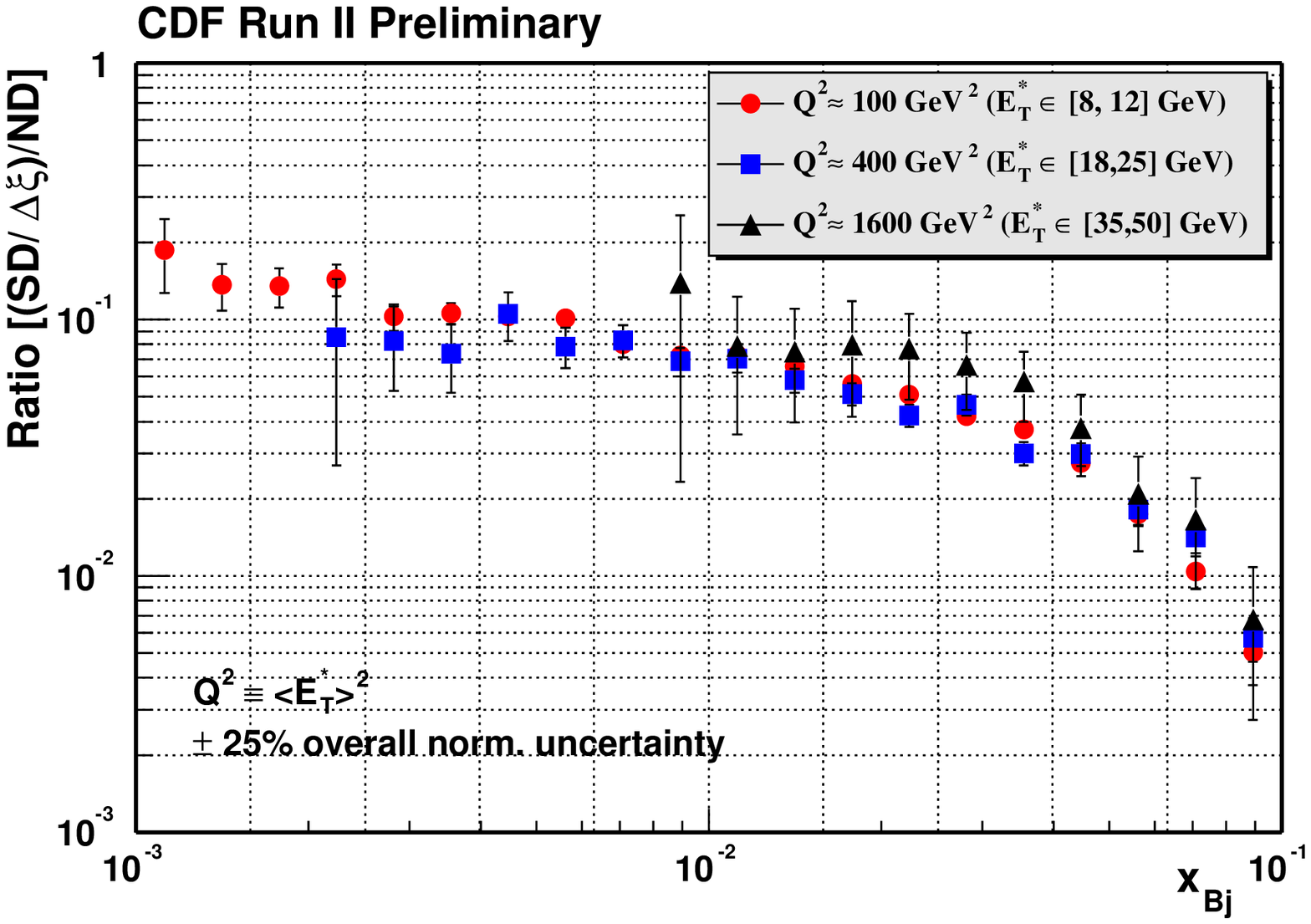}
\caption{Ratio of diffractive to non-diffractive di-jet event rates as a function of $x_{B_j}$
at CDF for different $\xi$ ranges and compared to Run I (a) and for different $Q^2$ values.}
\label{fig:SD}
\end{figure}

One of the central issues of diffraction is whether hard diffractive
processes obey QCD factorization. As Collins proved
\cite{Collins:1997sr} for the general class of diffractive DIS processes, the cross
section can be described as the convolution of a hard scattering
matrix element (process dependent) and parton density functions
(process independent). The question arises if this factorization
theorem is more general, are parton densities really universal in
diffractive exchange? Can we use them for different collider processes
and energies?  To answer this question is fundamental for the
understanding of diffraction and it is also important to
extrapolate Tevatron results to the LHC. The general strategy to
prove factorization is to extract parton density functions (PDFs) and
compare predictions to measurements of other processes and
experiments.

Before the extraction of PDFs, diffractive fractions already yield some
insight. Both, CDF and \D0 , measured for different processes the
fractions of events with one gap to all events. It was found that all
ratios are at the order of 1\% at the Tevatron which would support a
more general factorization theorem. However the ratio of 1\% yields an
uniform gap suppression w.r.t. HERA where the diffractive rates are
approximately 10 times higher. This discrepancy indicates already the
breakdown of QCD factorization.

CDF measured in Run I the diffractive structure function of the
anti-proton from SD di-jets and the result was compared with
expectations from diffractive DIS measurements of H1
\cite{Affolder:2000vb}. These events can be described in terms of a
Pomeron emitted from the anti-proton and scattering with a parton from
the proton.
%$\xi$ describes the fraction of momentum loss of the proton and $\beta$ 
%is the fraction of Pomeron momentum carried by the interacting parton.
The diffractive structure function was obtained by measuring the ratio
of the diffractive and non-diffractive cross section. The product of
this ratio with the known non-diffractive structure function gives the
result on the diffractive structure function. Although the shapes of
the structure functions from HERA and Tevatron have similar shapes,
there is a normalization discrepancy of a factor of 10. This result
again confirms the breakdown of QCD factorization between the Tevatron
and HERA as already expected from the result on the diffractive
fractions. One possibility to explain these observations is that,
since there are more spectator partons in $p\bar p$ collisions
w.r.t. $\gamma^* p$ collisions, the rate of gap destructions due to
soft partonic interactions is larger at the Tevatron. One attempt to
describe this rate is made with the introduction of the concept of the
``gap survival probability'' $|S|^2$ \cite{Bjorken:1992er,
Gotsman:1993vd}.  The observation that the shapes of the two structure
functions from HERA and the Tevatron are similar supports this
concept. The gap survival probability factor corrects the
normalization discrepancy.

In Run II the diffractive di-jet sample was collected with a dedicated
trigger which selects events with at least one calorimeter tower above
the 5 GeV $E_{\rm T}$ threshold and a threefold Roman Pot Spectrometer
coincidence. Calorimeter information is used to determine the momentum
loss of the anti-proton,
\begin{equation}
\xi_{\bar p}=\frac{\sum E_T e^{-\eta}}{\sqrt s} \, .
\end{equation}
SD and background regions are selected according to the measured
$\xi_{\bar p}$ values. A large number of events are at $\xi_{\bar
p}\sim 1$ which are due to the overlap of at least one ND
contribution. A plateau is observed in the $\xi_{\bar p}$ distribution
which results from a distribution proportional to $1/\xi_{\bar p}$ as
expected for diffractive production.

\begin{figure}[t]\hspace{-.5cm}
(a)\includegraphics[width=.55\textwidth]{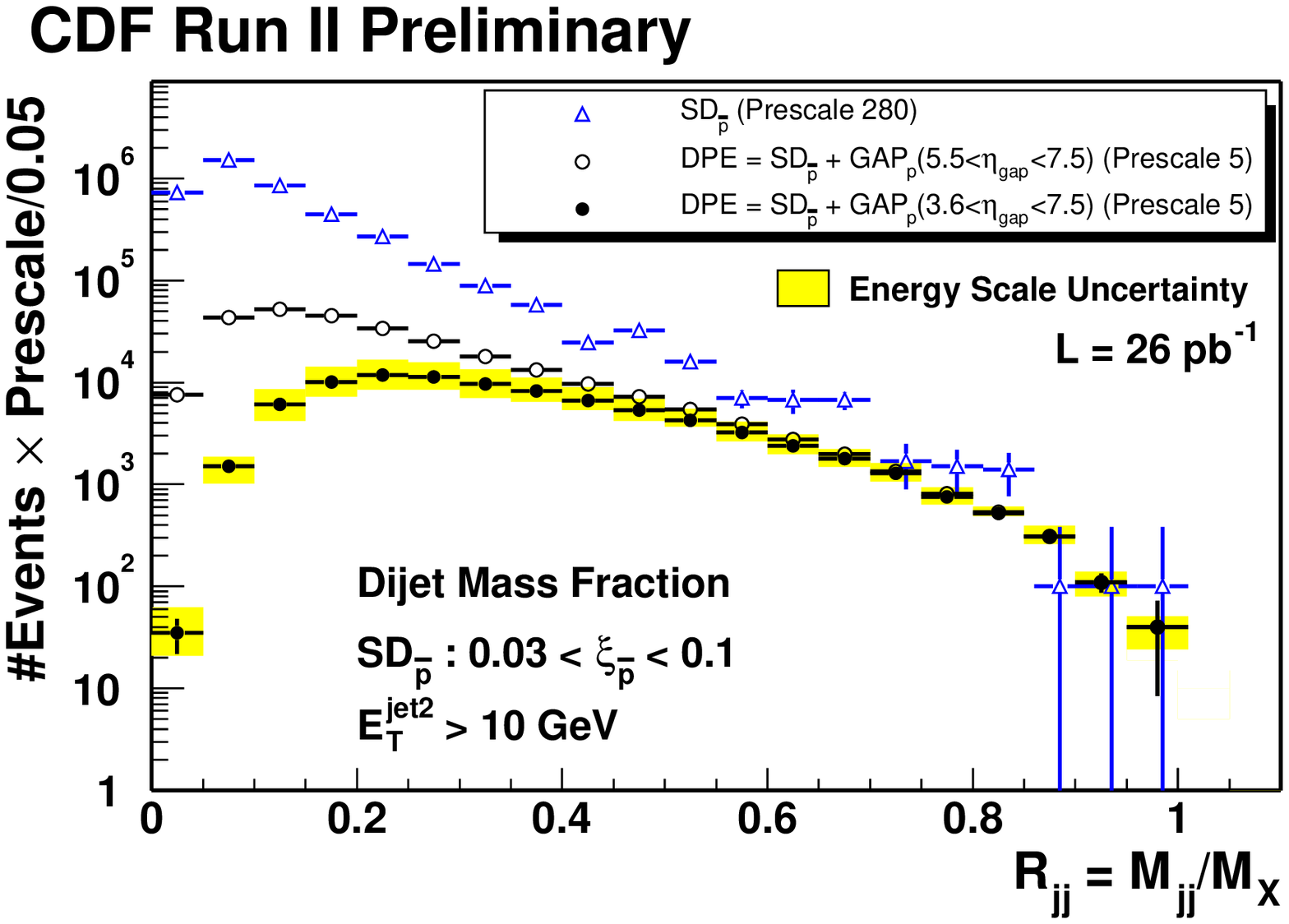}
(b)\includegraphics[width=.45\textwidth]{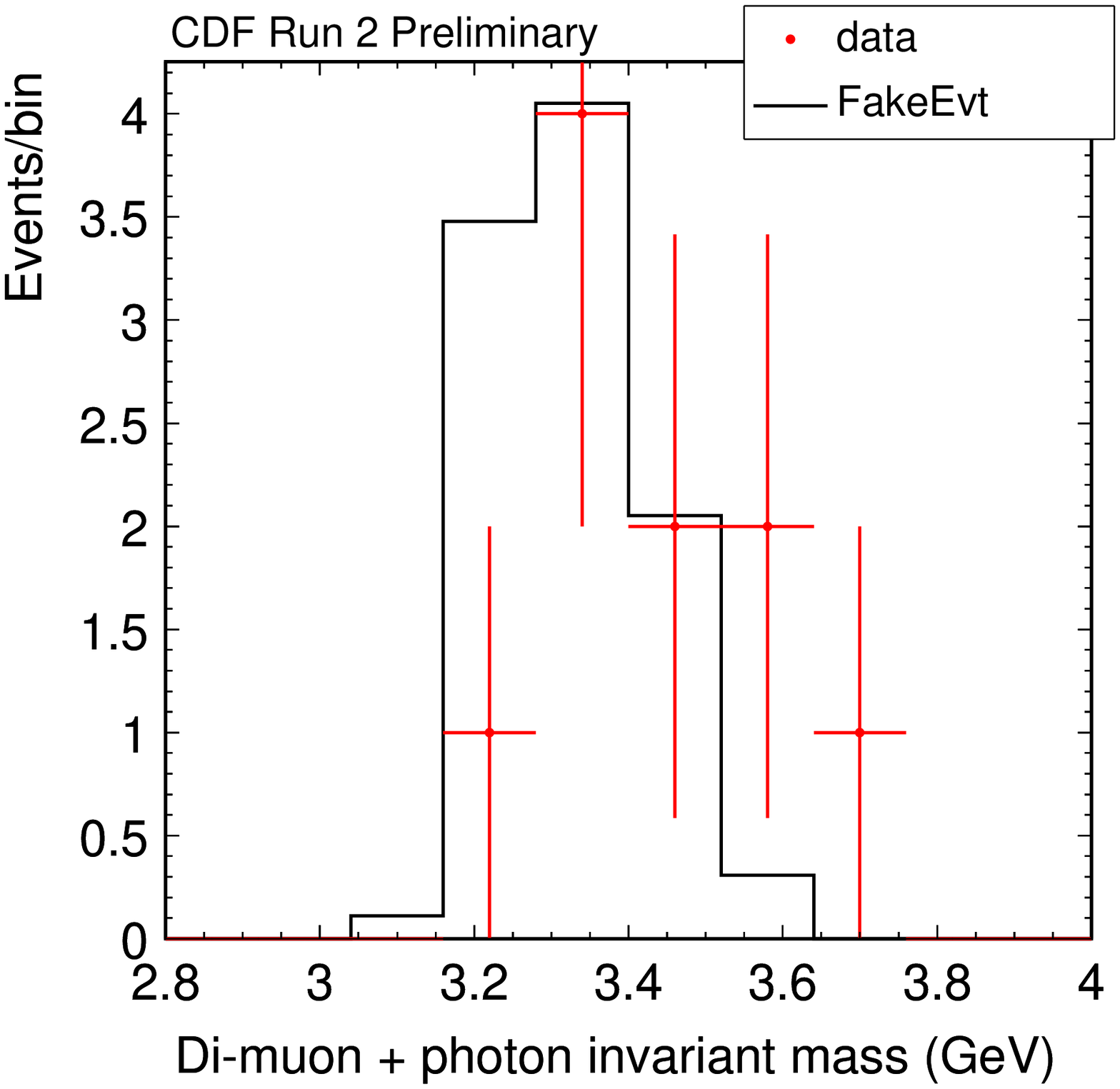}
\caption{Di-jet mass fraction for different rapidity gap selections (a) 
and di-muon plus photon invariant mass in the exclusive sample 
compared to Monte Carlo predictions (b) at CDF.}
\label{fig:DPE}
\end{figure}

In Fig.~\ref{fig:SD} the ratio of SD to ND event rates is plotted
versus Bjorken $x_{B_j}$. In Fig.~\ref{fig:SD}(a) this ratio is
integrated over three different $\xi$ regions and compared to the Run
I result.  There is no $\xi$ dependence observable between 0.03 and
0.1. Furthermore the slope and normalization agrees and thus confirms
the Run I result. In Fig.~\ref{fig:SD}(b) the same ratio is plotted
for three different $Q^2$ values, where $Q^2$ is the mean jet
transverse energy $\langle (E^1_{\rm T} + E^2_{\rm
T})/2\rangle^2$. There is no appreciable $Q^2$ dependence in the
observable region of $100 < Q^2 < 1600$ GeV$^2$. Thus both structure
functions seem to have a similar $Q^2$ evolution which does not
disfavor any of the two mechanisms of hard diffraction, namely the
existence of a hard Pomeron (exchange of a colorless object) or a soft
color rearrangement in the final state.

Using the di-jet sample CDF extracted also DPE di-jet production
events. These events are characterized by a leading anti-proton, two
jets in the central pseudorapidity region and a large rapidity gap on
the outgoing proton side. If factorization holds the ratio of DPE to
SD, $R^{DPE}_{SD}(x_p, \xi_p)$ has to be equal to the ratio of SD to ND,
$R^{SD}_{ND}(x_{\bar p}, \xi_{\bar p})$ (in LO QCD) for a fixed $x_{B_j}$
and $\xi$ value.  Although the collected events have different $\xi$
ranges for the proton and anti-proton, the weighted average of
$R^{SD}_{ND}(x_{\bar p}, \xi_{\bar p})$ is flat in $\xi$ and the ratio
was extrapolated to $\xi = 0.02$. At this $\xi$ values the above
mentioned ratios differ by a factor of 5 \cite{Affolder:2000hd}. The
deviation from unity yields again a breakdown of factorization. Since
the formation of a second gap is less suppressed this result is
coherent with the concept of the gap survival probability. The number
of spectator partons does not changed with the formation of a second
gap, i.e. one does not have to pay the price for the gap two times.

In the same manner as was previously done by the SD/ND ratio, the
diffractive structure function can be extracted from the DPE/SD
ratio. The obtained result now approximately agrees with expectations
from H1 leading again to the above mentioned conclusions. 

\section{Search for exclusive events}

Since the CDF Roman Pots have been installed at the end of Run I, CDF
collected in Run II two orders of magnitude more di-jet data which made
a study of exclusive di-jet production in DPE possible. The strategy
was to obtain an inclusive DPE di-jet event sample and look for
exclusive signature using the di-jet mass fraction $R_{jj}=M_{jj}/M_X$
where the di-jet mass is divided by the mass of the rest of the system
excluding the proton and anti-proton. In Fig.~\ref{fig:DPE}(a) the
obtained number of events is plotted versus the di-jet mass fraction
where no gap, a narrow gap and a wide gap was required on the proton
side. The result is a smoothly falling spectrum all the way down to
$R_{jj}=1$ and a similar event yield at high mass fraction regardless
of the gap requirements. From this it follows that no significant excess due to
exclusive di-jets is seen at high $R_{jj}$. An upper limit on the
exclusive di-jet production cross section is calculated based on events
with $R_{jj} > 0.8$. For example requiring a minimum jet transverse energy of 10
GeV, this upper limit is 
\begin{equation}
E_{\rm T}^{\rm min}=10 ~{\rm GeV}:~~\sigma (R_{jj}>0.8)<1.1\pm 0.1(stat)\pm 0.5 (sys)~~ {\rm nb} \, .
\end{equation}

There are also other production channels available in DPE, we mention
here the exclusive $\chi_c^0$ production. This process is of
particular interest, since the $\chi_c^0$ quantum numbers are very
similar to the ones of the Higgs boson, thus it can be used to test
and normalize the predictions for exclusive Higgs production at the
LHC. CDF did an analysis in Run II where the $\chi_c^0$ further decays
in a $J/\psi +\gamma$. 93 pb$^{-1}$ of di-muon triggered data was used
and events have been selected in the $J/\psi$ mass window. A
large rapidity gap on both the proton and anti-proton side was
required. 10 events have been found which are exclusive $\chi_c^0 (\to
J/\psi +\gamma)$ candidates. In Fig.~\ref{fig:DPE}(b) the di-muon plus
photon invariant mass of the 10 events is compared with a sample of
generated $\chi_c^0$ events passed through a detector simulation. The
invariant mass is consistent with that of the $\chi_c^0$, although the
mean mass is higher and the distribution broader in the data than in
the simulation. This may be due to the simulation or there may be
contributions from cosmic events, higher mass $\chi_c$ mesons or
$J/\psi + \pi^0$ events. Since it is very difficult to fully
understand the background one may calculate an ``upper limit'' on
exclusive $\chi_c^0$ production assuming that the observed 10 events
are all $J/\psi +\gamma$ events. This upper limit is:
\begin{equation}
\sigma = 49 \pm 18 (stat) \pm 39(sys)~~ {\rm pb} \, .
\end{equation}

\section{\D0 ~Forward Proton Detectors}

\begin{figure}[t]
\includegraphics[width=1\textwidth]{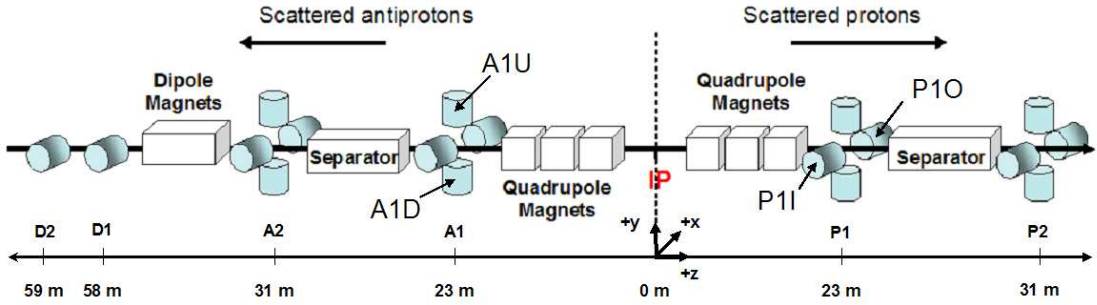}
\caption{The Forward Proton Detector at \D0 . Quadrupole Pots are named P or A 
when placed on the proton or the anti-proton side, respectively. Dipole Pots are named D.}
\label{fig:fpd}
\end{figure}

The \D0 ~Forward Proton Spectrometers have been installed and recently
commissioned. These are 9 momentum spectrometers with 2 scintillating
fiber detectors each, as schematically shown in
Fig.~\ref{fig:fpd}. There is a dipole spectrometer located on the
scattered anti-proton side behind the dipole magnets approximately 58 m
away from the interaction point. They have in the range of $|t|\approx
0 - 1$ GeV$^2$ and $\xi\approx 0.03 - 0.07$ a very good
acceptance. Eight quadrupole spectrometers are on both side of the
main detector approximately 23 and 31 m away from the interaction
point, behind the quadrupole magnets and the separators. They have a
very good acceptance in the region of: $|t|\approx 0.8 - 3.0$ GeV$^2$
and $\xi\approx 10^{-3} - 0.05$. The position detectors housed inside
Roman Pots operate millimeters away from the beam, however outside of the
ultra high vacuum. They enable the reconstruction of the high energy
protons and anti-protons directly, thus providing a first time
possibility of tagging both the protons and anti-protons and measuring
their $\xi$ and $t$ dependence at the Tevatron.

The \D0 ~scintillating fiber detectors have 6 layers of scintillating
fiber channels and one trigger scintillator layer. The fibers are
oriented within $\pm 45^o$ to reconstruct hits and obtain
redundancy. Furthermore every second channel is offset by 2/3 fiber
for a finer hit resolution. All 18 detectors regularly brought close
to the beamline and diffractive samples being collected.

\begin{figure}
\includegraphics[width=.6\textwidth]{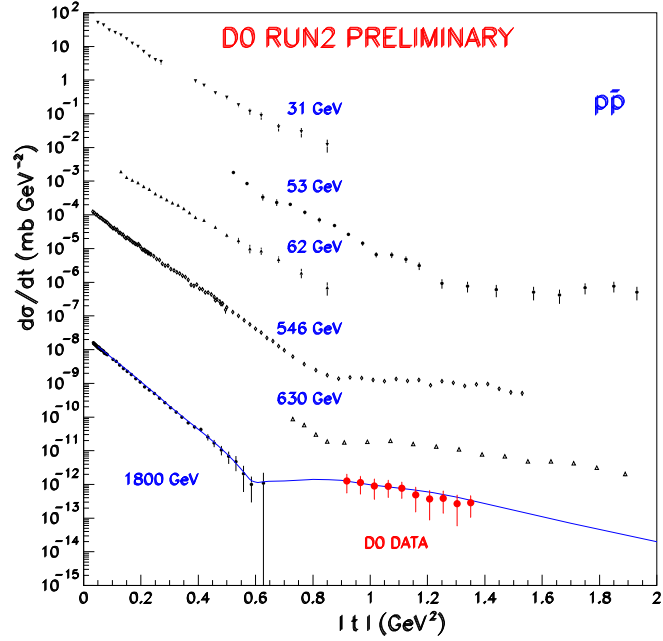}
\caption{The slope of the elastic $p\bar p$ cross section measured at \D0 ~and compared to 
predictions of \cite{Block:1989gz} (solid line).}
\label{fig:elastic}
\end{figure}

In a small dedicated test run of the FPDs in a stand alone mode the
slope of the elastic cross section of proton anti-proton scattering was
measured. In Fig.~\ref{fig:elastic} the result (not normalized) is plotted and compared
with theory predictions \cite{Block:1989gz}. \D0 ~access a new kinematic domain
with these measurements and the result of the slope agrees well with
the model of Bock {\it et al.} \cite{Block:1989gz}. This measurement is being
redone using the fully integrated FPDs. The alignment and detector
understanding of the Roman Pot detectors is in progress and more
physics results are expected soon.

%%%%%%%%%%%%%%%%%%%%%%%%%%%%%%%%%%%%%%%%%%%%%%%%
%% BACKMATTER
%%%%%%%%%%%%%%%%%%%%%%%%%%%%%%%%%%%%%%%%%%%%%%%%

%\begin{theacknowledgments}
%\end{theacknowledgments}



\bibliographystyle{aipproc}   % if natbib is available

\bibliography{k_peters}

\IfFileExists{\jobname.bbl}{}
 {\typeout{} \typeout{******************************************}
 \typeout{** Please run "bibtex \jobname" to obtain} \typeout{** the
 bibliography and then re-run LaTeX} \typeout{** twice to fix the
 references!}  \typeout{******************************************}
 \typeout{} }

\end{document}